\documentclass[aps,pra,english,twocolumn,showpacs,preprintnumbers,amsmath,amssymb,floatfix,superscriptaddress]{revtex4-1}
\usepackage{graphicx}
\usepackage{amssymb}
\usepackage{babel}
\makeatletter
\usepackage{multirow}

\usepackage[version=3]{mhchem}
\newcommand{\mi}{\mathrm{i}}
\newcommand{\me}{\mathrm{e}}
\usepackage{color}

\usepackage{ulem}

\makeatother
\usepackage{babel}
\makeatother

\begin{document}

\title{Impact of effective polarisability models on the predicted release dynamics of CH$_4$ and CO$_2$ from premelted ice}

\author{J. Fiedler}
\email{johannes.fiedler@physik.uni-freiburg.de}
\affiliation{Physikalisches Institut, Albert-Ludwigs-Universit{\"a}t Freiburg, Hermann-Herder-Str. 3, D-79104 Freiburg, Germany}
\affiliation{Centre for Materials Science and Nanotechnology,
Department of Physics, University of Oslo, P. O. Box 1048 Blindern,
NO-0316 Oslo, Norway}

\author{P. Thiyam}
\email{thiyam@kth.se}
\affiliation{Department of Materials Science and Engineering, Royal Institute of Technology, SE-100 44 Stockholm, Sweden}
\affiliation{Department of Energy and Process Engineering, Norwegian University of Science and Technology, NO-7491 Trondheim, Norway}

\author{F. A. Burger}
\affiliation{Physikalisches Institut, Albert-Ludwigs-Universit{\"a}t Freiburg, Hermann-Herder-Str. 3, D-79104 Freiburg, Germany}

\author{D. F. Parsons}
\affiliation{Department of Applied Mathematics, Australian National University, Canberra, Australia}

\author{M. Walter}
\affiliation{Functional Nanosystems, Albert-Ludwigs-University Freiburg, Hermann-Herder-Str. 3, D-79104 Freiburg, Germany}

\author{I. Brevik}
\affiliation{Department of Energy and Process Engineering, Norwegian University of Science and Technology, NO-7491 Trondheim, Norway}

\author{C. Persson}
\affiliation{Centre for Materials Science and Nanotechnology, Department of Physics, University of Oslo, P. O. Box 1048 Blindern, NO-0316 Oslo, Norway}
\affiliation{Department of Materials Science and Engineering, Royal Institute of Technology, SE-100 44 Stockholm, Sweden}

\author{S. Y. Buhmann}
\email{stefan.buhmann@physik.uni-freiburg.de}
\affiliation{Physikalisches Institut, Albert-Ludwigs-Universit{\"a}t Freiburg, Hermann-Herder-Str. 3, D-79104 Freiburg, Germany}
\affiliation{Freiburg Institute for Advanced Studies, Albert-Ludwigs-Universit\"at Freiburg, Albertstr. 19, D-79104 Freiburg, Germany}

\author{M. Bostr{\"o}m}
\email{Mathias.A.Bostrom@ntnu.no}
\affiliation{Department of Energy and Process Engineering, Norwegian University of Science and Technology, NO-7491 Trondheim, Norway}

\begin{abstract}
We present a theory for Casimir--Polder forces acting on greenhouse gas molecules dissolved in a thin water film. Such a nanosized film has recently been predicted to arise on th surface of melting ice as stabilized by repulsive Lifshitz forces.  We show that different models for the effective polarizability of greenhouse gas molecules in water lead to different predictions for  how Casimir--Polder forces influence the extraction of CH$_4$ and CO$_2$  molecules from the melting ice surface. In the most intricate model of a finite-sized molecule inside a cavity, dispersion potentials push the methane molecules towards the ice surface whereas the carbon dioxide typically will be attracted towards the closest interface (ice or air).  Previous models for effective polarizability had suggested that CO$_2$ would also be pushed towards the ice surface. Release of greenhouse gas molecules from the surface of melting ice can potentially influence climate greenhouse effects. 
 \end{abstract}

\maketitle

\section{Introduction}

 Theoretical and experimental interest has been directed towards an understanding of fluctuation induced dispersion forces (Casimir, Casimir--Polder and van der Waals forces) in  the last
decades\,\cite{Pars,Ninhb,Mahanty2,Buhmann12a}.
In the process  of ice melting  at the triple point a nanosized  film of
water  has been predicted. It is stabilized by repulsive Casimir forces\,\cite{Elbaum,Elbaum2,Wilen}. 
One important question is how Casimir--Polder forces  influence release of
greenhouse gas molecules trapped from surface of porous ice.  
Some experiments indicate enhanced concentrations of methane in drinking water associated with shale-gas extraction\,\cite{Osborn,Vidic}.
\begin{figure}[h!]
 \centering
 \includegraphics[width=0.6\columnwidth]{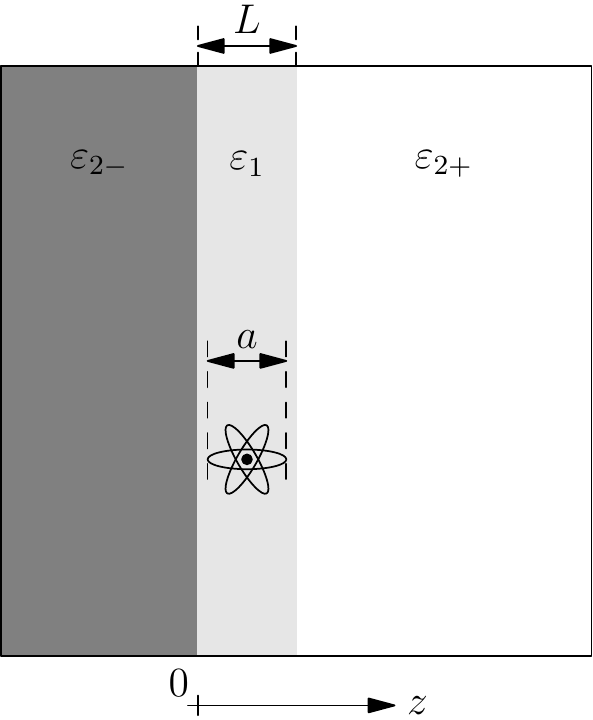}
 \caption{(Color online) Sketch of the considered arrangement: a particle of radius $a$ is embedded in a medium, $\varepsilon_1$ of thickness $L$, which is planarly in contact with two other media $\varepsilon_{2\pm}$. The particle's position is denoted by $z$ and is with respect to the left interface of the system. We consider an ice-water-air system.}\label{fig:arra}
\end{figure}
  In a similar way, release of methane  from melting ice  in the arctic due to increased temperatures may enhance climate greenhouse effects and pose threats to wildlife\,\cite{Whiteman,Kort}.
We explore in this article how molecules dissolved in the equilibrium ice-water-air
system experience Casimir--Polder forces. We focus our
attention on methane and carbon dioxide molecules in the water phase. The Casimir--Polder force  influences
the extraction of greenhouse gas molecules, starting from being near the ice surface, then going into the bulk of the water nanosheet on the
ice surface, and finally towards the water-air interface. New models for the effective polarizability of greenhouse gas molecules in water, which account for the finite size of the particles and the vacuum bubble arising from Pauli repulsion  will be used\,\cite{JohannesJCPA2017}.

Notably, the most advanced model predicts (in contrast to simpler models) Casimir--Polder forces to push carbon dioxide molecules towards the closest interface. In simpler models previously used the carbon dioxide is pushed towards the ice surface. In contrast, methane is in all considered models pushed towards the ice surface. This means that carbon dioxide and methane will behave fundamentally different inside a premelting water layer. We will calculate the Casimir--Polder potential acting on the molecules inside the three-layer system and show that both will behave differently at the water-air interface by a change from attractive to repulsive force.

\section{Theory}

We have recently demonstrated that finite size effects acting on atoms and molecules can be neglected at distances larger than close contact \cite{PriyaPRE14}. We therefore neglect finite size effects and use the local-field corrected van der Waals force acting on polarizable particles (e.g. CH$_4$ and CO$_2 $  molecules) in a three layer system (ice-water-air, see Fig.~\ref{fig:arra}).  The  thickness ($L=32$ \AA) of the water nanosheet at the triple point with ice-water-air  at equilibrium is given by Lifshitz forces acting on the system\,\cite{Elbaum}.

The Casimir--Polder potential for a particle embedded between two infinite half spaces separated by the distance $L$ can be written as \cite{Buhmann12a}
\begin{eqnarray}
 \lefteqn{U(z) = \frac{ \mu_0 k_B T}{4\pi}\sum_{n=0}^\infty{}'  \xi_n^2 \alpha^\star(\mi\xi_n) \int\limits_0^\infty \mathrm dk^\parallel\frac{k^\parallel}{\kappa_1^\perp}}\nonumber\\
 &&\times\left\lbrace \left[\frac{r_s^{-}}{D_s} - \left(1+2\frac{{k^\parallel}^2 c^2}{\varepsilon_1(i\xi_n)\xi_n^2}\right)\frac{r_p^{-}}{D_p}\right] \me^{-2\kappa_1^\perp z}\right.\nonumber \\&&
 \left. + \left[\frac{r_s^{+}}{D_s}-\left(1+2\frac{{k^\parallel}^2 c^2}{\varepsilon_1(i\xi_n)\xi_n^2}\right)\frac{r_p^{+}}{D_p}\right]\me^{-2 \kappa_1^\perp(L-z)}\right\rbrace \,,\label{eq:CP}
\end{eqnarray}
where $k_B$ is the Boltzmann constant, $T$ is the temperature, and the prime indicates that the $n=0$ term should be divided by 2. The corresponding Fresnel reflection coefficients for $s$- and $p$-polarized light read
\begin{eqnarray}
 r_s^{\pm} &=& \frac{\kappa_1^\perp-\kappa_{2\pm}^\perp}{\kappa_1^\perp+\kappa_{2\pm}^\perp}\, ,\quad r_p^\pm=\frac{\varepsilon_{2\pm} \kappa_1^\perp-\varepsilon_1 \kappa_{2\pm}^\perp}{\varepsilon_{2\pm} \kappa_1^\perp+\varepsilon_1 \kappa_{2\pm}^\perp}\, ,
\end{eqnarray}
with the imaginary part of the perpendicular component of the wave vector
\begin{equation}
 \kappa_j^\perp = \sqrt{\varepsilon_j(i\xi)\frac{\xi^2}{c^2}+{k^\parallel}^2} \, , 
\end{equation}
and the multiple reflection coefficients
\begin{equation}
 D_{s,p} =1-r_{s,p}^+ r_{s,p}^- \mathrm e^{-2\kappa_1^\perp L} \, .
\end{equation}
In the non-retarded limit ($k^\parallel \gg \sqrt{\varepsilon_j} \omega/c $) the perpendicular component of the wave vectors simplify to
\begin{equation}
\kappa^\perp_1 = \kappa^\perp_2 = \kappa^\perp_3 \approx k^\parallel \, ,
\end{equation}
leading to vanishing reflection coefficients for $s$-polarized waves, $r_s^\pm=0$. For $p$-polarized waves they simplify to the Fresnel reflection coefficients
\begin{equation}
 r_p^\pm = \frac{\varepsilon_{2\pm}-\varepsilon_1}{\varepsilon_{2\pm}+\varepsilon_1} \, .
\end{equation}
Thus, the integral along the $k$-axis can be performed and the Casimir--Polder potential in the nonretarded limit, Eq.~(\ref{eq:CP}), results in
\begin{equation}
 U(z) = -\frac{C_3^-(z)}{z^3} - \frac{C_3^+(z)}{(L-z)^3} \, , \label{eq:addC}
\end{equation}
with the distance-dependent $C_3^\pm(z)$ coefficients
\begin{figure}[t]\vspace{0.5cm}
 \includegraphics[width=0.9\columnwidth]{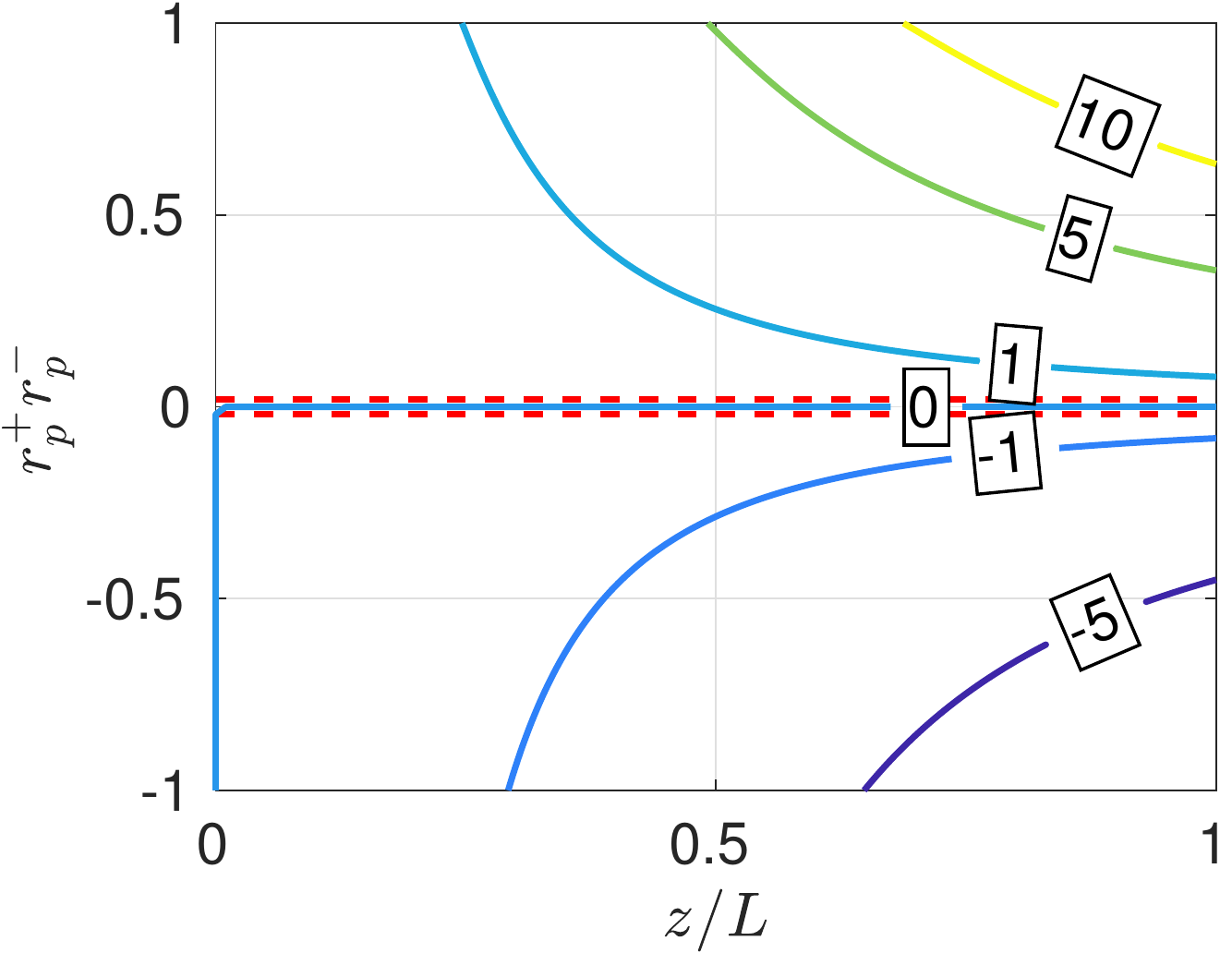}
 \caption{Relative deviation of the $C^-_3$ coefficient in percent caused by multiple reflections in the planar cavity depending on the position of the particle and the reflectivity at the outer interfaces. The frequency dependence is included in the reflection coefficient $r_p^\pm(\omega)$. The dashed red lines mark the relevant range for the considered  scenario in this article. It can be observed that the influence of multiple reflecting occurs close to the second surface and gets stronger by the transition to perfectly reflecting plates.}
 \label{fig:multi}
\end{figure}
\begin{widetext}
\begin{eqnarray}
 C_3^- &=& \frac{ k_B T}{8\pi\varepsilon_0}\sum_{n=0}^\infty{}'   \frac{\alpha^\star(\mi\xi_n)}{\varepsilon_1(\mi\xi_n)} r_p^- {}_4F_3\left(1,\frac{z}{L},\frac{z}{L},\frac{z}{L};1+\frac{z}{L},1+\frac{z}{L},1+\frac{z}{L};r_p^+r_p^- \right)\,,\\
C_3^+ &=&\frac{ k_B T}{8\pi\varepsilon_0}\sum_{n=0}^\infty{}'   \frac{\alpha^\star(\mi\xi_n)}{\varepsilon_1(\mi\xi_n)} r_p^+{}_4F_3\left(1,\frac{L-z}{L},\frac{L-z}{L},\frac{L-z}{L};\frac{2L-z}{L},\frac{2L-z}{L},\frac{2L-z}{L};r_p^+r_p^- \right)\,.
\end{eqnarray} 
\end{widetext}
The generalized hypergeometric function ${}_4F_3$ denotes a measure for the impact of multiple reflections on the Casimir--Polder potential. Figure~\ref{fig:multi} illustrates the behavior of this function and marks the relevant parameter regime for the scenario considered here. It can be seen that the force is additive with respect to both interfaces for this situation. This is caused by the weak reflectivity of the water-ice interface, the dielectric functions of water and ice being similar.

By neglecting multiple reflections in the given geometry we can conclude that retardation effects can be neglected as well. Retardation plays an important role at larger distances. A well reflecting cavity, as is formed by the middle layer bound by two interfaces, virtually increases the path length of the propagating waves. Thus, retardation will be important for cavities with high reflection coefficients. As this is not the case in the considered scenario, these effects will not be important. However, we will compare the approximated potentials with the exact one.

\begin{figure}[t]
 \centering
 \includegraphics[width=\columnwidth]{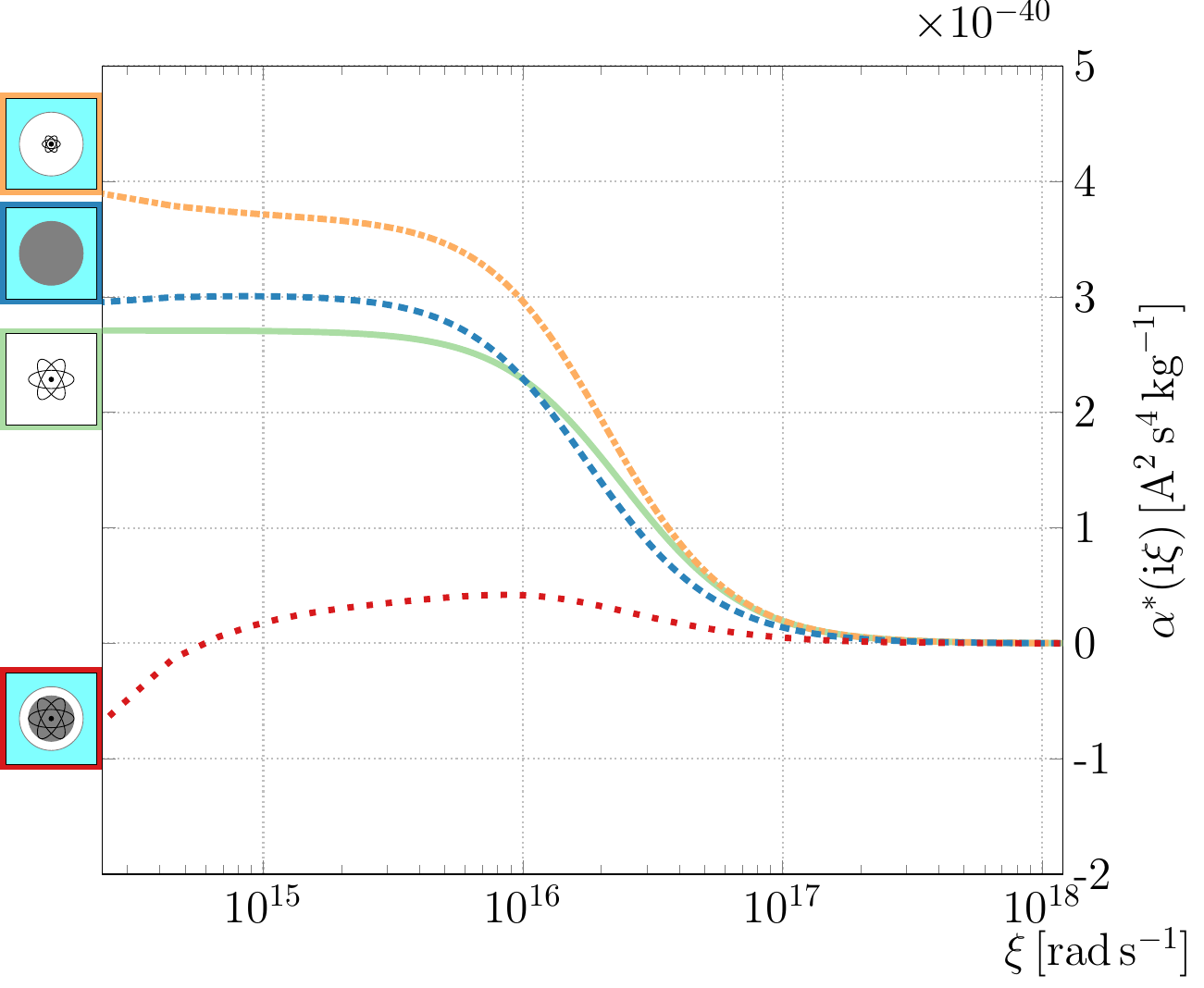}
 \caption{(Color online) Graph of the frequency dependence of polarizability for methane (solid green line) and the corresponding effective polarizabilities via the hard-sphere model, Eq.~(\ref{Eq2d}), (dashed blue line), Onsager's real cavity model, Eq.~(\ref{Eq2a}), (dashed-dotted orange  line) and the finite-size model, Eq.~(\ref{Eq2e}), (dotted red line).}\label{fig:CH4}
\end{figure}

Furthermore $\alpha^\star(\mi\xi_n )$ is the effective molecular polarizability in water at the Matsubara frequencies  $\xi_n=2 \pi k_B T n/\hbar$\,\cite{Sernelius}. For the free-space polarizabilities the adjusted parameters $\alpha_j$ and ionization potentials $\omega_j$ given in Ref.~\cite{JohannesJCPA2017} were fitted to agree with the free space polarizability obtained from ab initio calculations\,\cite{ParsonsNinham2010dynpol}
\begin{equation}
\alpha(i \xi_n)=\sum_j\frac{\alpha_j}{1+(\xi_n/\omega_j)^2} \, .
\label{Eq2b}
\end{equation}
These and the hard sphere radii $a$ for methane and carbon dioxide were derived as in several papers by Parsons and Ninham\,\cite{ParsonsNinham2009,ParsonsNinham2010dynpol}. The cavity radii $a_C$ were derived by solving the electrostatic Maxwell equations for particles embedded in a continuous medium. The different radii were taken from Ref.~\cite{JohannesJCPA2017}.
The combination of pressure and temperature at which liquid water, solid ice, and water vapor can coexist in a stable equilibrium occurs at $T=273.16 K$  and a partial vapor pressure of 611.73 Pa.
The dielectric functions of ice ($\varepsilon_{2-}$), water  ($\varepsilon_1$), and vapor (air) ($\varepsilon_{2+}=1$) were taken from the work of Elbaum and Schick\,\cite{Elbaum}.

Three different models are used for the effective polarizabilities
$\alpha^\star(i \xi_n)$ of the molecule in water:
(i)~Onsager's real-cavity model  for local-field corrections assumes
that the molecule is situated in a small spherical vacuum bubble
embedded in the water medium\,\cite{Onsager}. One finds that
\cite{Sambale07,Sambale09}
%
\begin{equation}
\alpha^\star_{Ons}=\alpha\biggl(\frac{3\varepsilon_1}{2\varepsilon_1+1}\biggr)^2 \, .
\label{Eq2a}
\end{equation}
 (ii)~The hard sphere model posits that
the molecule can be described as a homogeneous dielectric sphere of radius $a$. Its
effective permittivity $\tilde\varepsilon$ can be deduced from the free-space
polarizability, Eq.~(\ref{Eq2b}) via the Clausius--Mossotti relation \cite{Jackson, JohannesJCPA2017}
%
\begin{equation}
\alpha=4\pi\varepsilon_0 a^3\,\frac{\tilde\varepsilon-1}{\tilde\varepsilon+2}\,.
\end{equation}
The excess polarizability of the homogeneous-sphere molecule in water
is then \cite{Sambale10}
%
\begin{equation}
\alpha^\star_{HS}=4\pi\varepsilon_0\varepsilon_1 a^3 \,\frac{\tilde\varepsilon-\varepsilon_1}
 {\tilde\varepsilon+2\varepsilon_1}\,.
\label{Eq2d}
\end{equation}
(iii)~A generalization of both models is a homogeneous-sphere
molecule embedded in a vacuum bubble of radius $a_C$ including finite-size effects of the particles (subscribe $fs$). The resulting
polarizability in water reads
%
\begin{equation}
\alpha^\star_{fs}=\alpha^\star_C+\alpha\,
 \biggl(\frac{3\varepsilon_1}{2\varepsilon_1+1}\biggr)^2
 \frac{1}{1+2 \alpha^\star_C\alpha/(8\pi^2\varepsilon_0^2\varepsilon_1a_C^6)}
\label{Eq2e}
\end{equation}
with
%
\begin{equation}
\alpha^\star_C=4\pi\varepsilon_0\varepsilon_1a_C^3\,\frac{1-\varepsilon_1}
 {1+2\varepsilon_1}
\end{equation}
denoting the excess polarizability of the bubble.

\begin{figure}[t]
 \centering
 \includegraphics[width=\columnwidth]{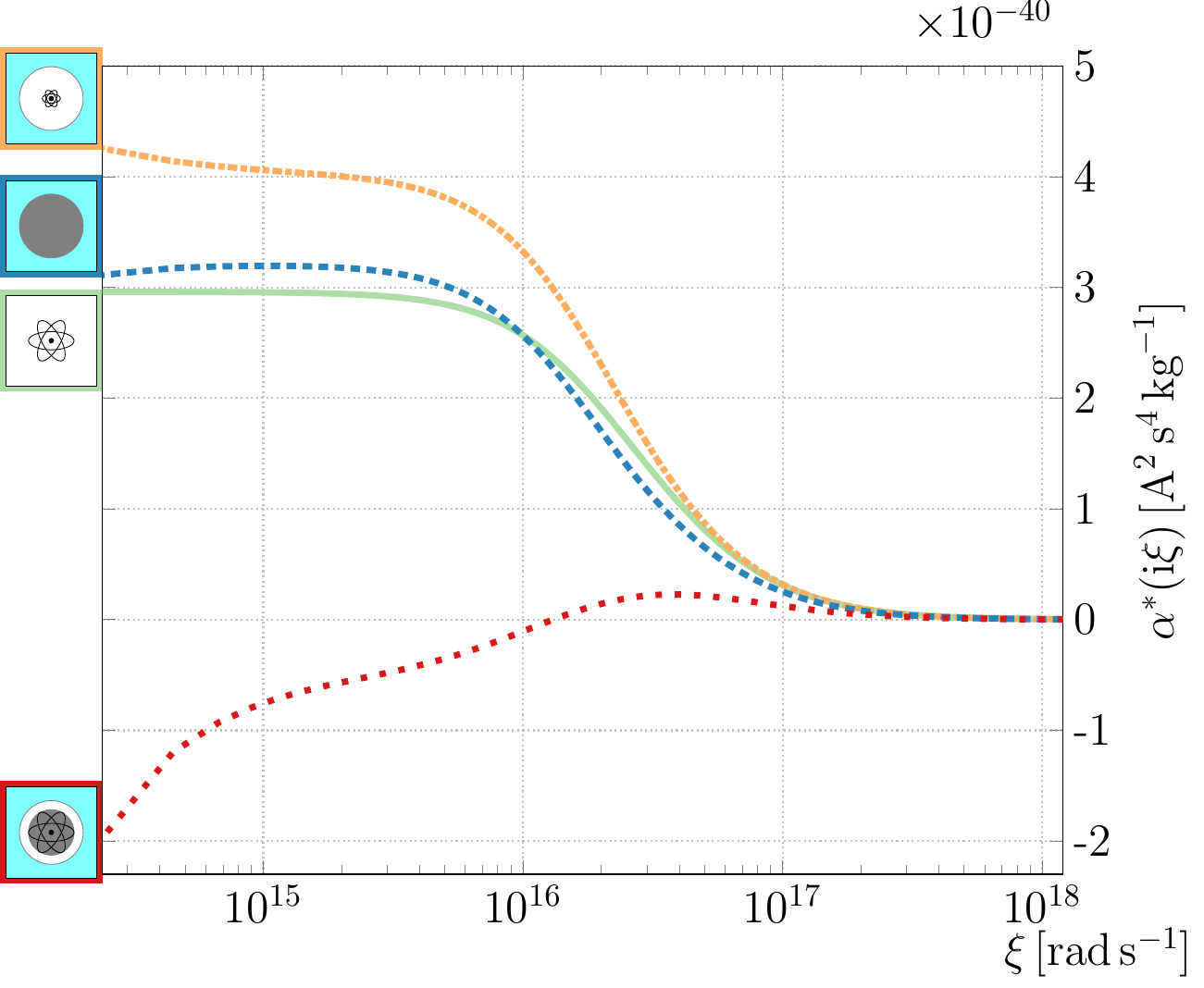}
 \caption{(Color online) Graph of the frequency dependence of polarizability for carbon dioxide (green solid line) and the corresponding effective polarizabilities via the hard-sphere model, Eq.~(\ref{Eq2d}), (blue dashed line), Onsager's real cavity model, Eq.~(\ref{Eq2a}), (orange dashed-dotted line) and the finite-size model, Eq.~(\ref{Eq2e}), (red dotted line).}\label{fig:CO2}
\end{figure}
All three models are based on physical assumptions with different depth of approximations. Onsager's real cavity assumes a point-like particle in a vacuum bubble and takes the transmission through the interface into account. This is valid for atoms in a large cavity. The hard-sphere model assumes a finite-size particle without a vacuum layer between the particle and the environmental medium and can be applied to larger particles such as clusters when the cavity radius is comparable to the particle's radius. The finite-size model is a composition of both models. It separates into two terms, the first one denotes the pure polarizability of an empty vacuum bubble and the second one is equivalent to Onsager's real cavity model and describes the transmission through the interface and the multiple scattering inside between the particle and the cavity. The finite-size model reduces to both other models by applying the corresponding limits. 

Applying the three models to methane and carbon dioxide embedded in water, one finds the effective polarizabilities depicted in Figs.~\ref{fig:CH4} and \ref{fig:CO2}, respectively. In both cases one finds an increase in the magnitude of the polarizability for the hard-sphere and Onsager's real cavity model compared to the free-space polarizabilities. One also observes that the finite-size model affects both cases very differently. For carbon dioxide, this model results in a large frequency region with negative polarizability, which is caused by the crossing of the dielectric functions of water and CO$_2$. In contrast, this model applied to methane shows a much smaller frequency region with negative values.
\begin{table}[htb]
 \begin{tabular}{c|l|cc}
  &Model & ${C}_3^- \ \text{in} \  \mu \mathrm{eV(nm)^3}$ & ${C}_3^+ \ \text{in} \  \mu \mathrm{eV(nm)^3}$\\\hline
  \parbox[t]{2mm}{\multirow{4}{*}{\rotatebox[origin=c]{90}{CH$_4$}}} 
& Free particle & 12.36  & -444.34 \\
 & Onsager's cavity & 10.86  & -553.96 \\
 & Hard-sphere & 5.92  & -406.42 \\
 & Finite-size & 3.20 & -45.87\\\hline 
  \parbox[t]{2mm}{\multirow{4}{*}{\rotatebox[origin=c]{90}{CO$_2$}}} 
& Free particle & 20.47 & -524.24 \\
 & Onsager's cavity & 19.22 & -651.27 \\
 & Hard-sphere & 13.91 & -483.68 \\
 & Finite-size & 9.44 & 31.26

 \end{tabular}
\caption{Table of $C_3$ coefficients for the water-ice interface (${C}_3^-$) and for the water-air interface (${C}_3^+$) for methane and carbon dioxide.} \label{tbl:C3M}
\end{table}
\section{Results}

\begin{figure}[b]
 \centering
 \includegraphics[width=\columnwidth]{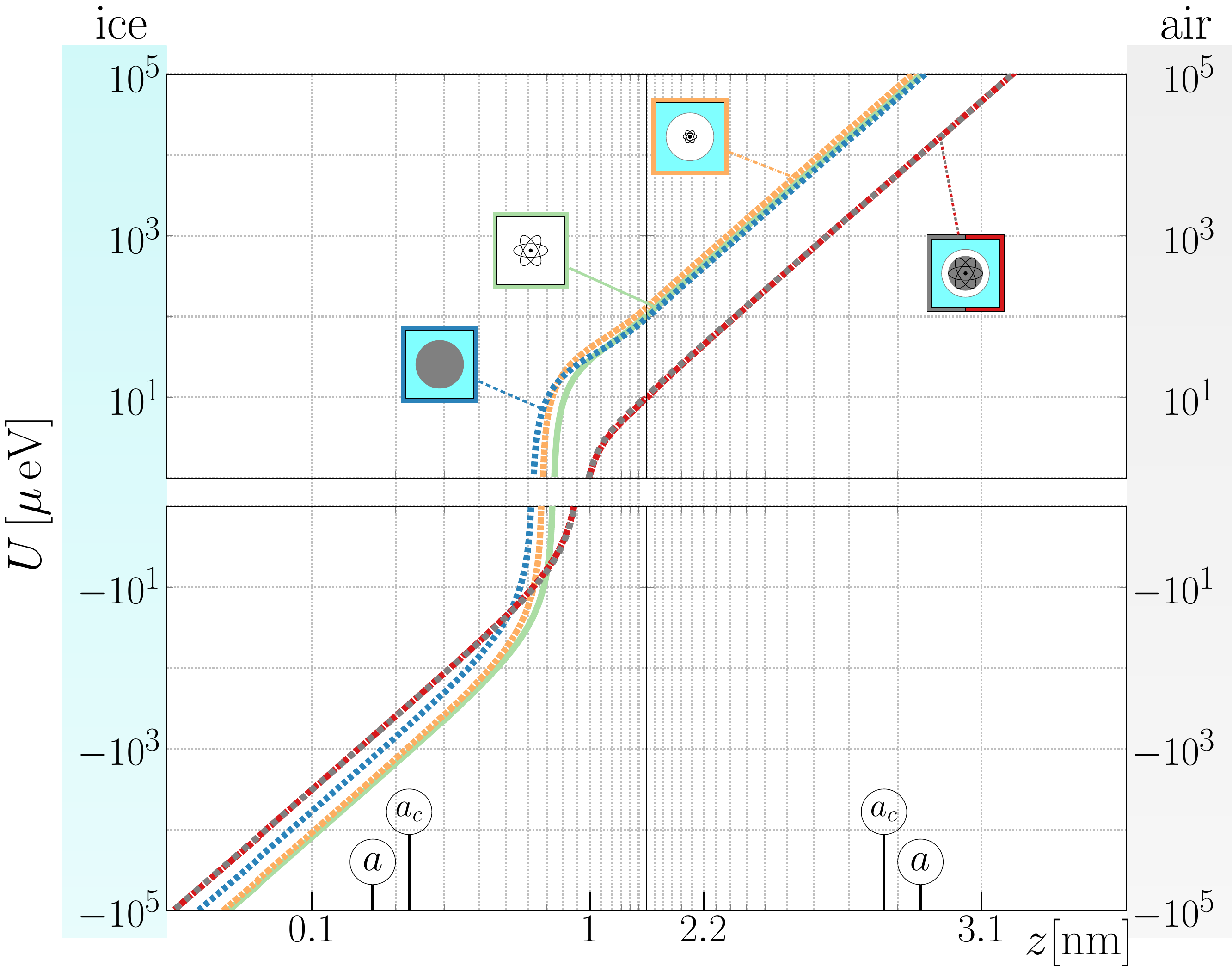}
 \caption{(Color online) Resulting Casimir--Polder potentials for methane using the free-space polarizability (green), hard-sphere model (blue), Onsager's real cavity model (orange), the finite-size model (red) and the nonretarded limit using the finite-size model (gray). The cavity radius $a_C$ and the particles radius $a$ are marked by needles on the $x$-axis to illustrate the physically possible positions of the particles.}\label{fig:pot2}
\end{figure}
The results for the $C_3$-coefficients of the water-air (right) and water-ice interface (left) are given in Table~\ref{tbl:C3M}. As predicted, the correction due to multiple reflections inside the cavity yields a relative error of less than 0.5 percent which is caused by the low reflection coefficient at the water-ice interface. Due to this fact, the resulting forces are independent of the thickness of the water layer. We restricted attention to the premelted water layer on top of an ice sheet; however the resulting Casimir--Polder potential is additive in the considered scenario with respect to both interfaces, see Eq.~(\ref{eq:addC}). The theory  is hence also applicable to thicker water layers.

\begin{figure}[t]
 \centering
 \includegraphics[width=\columnwidth]{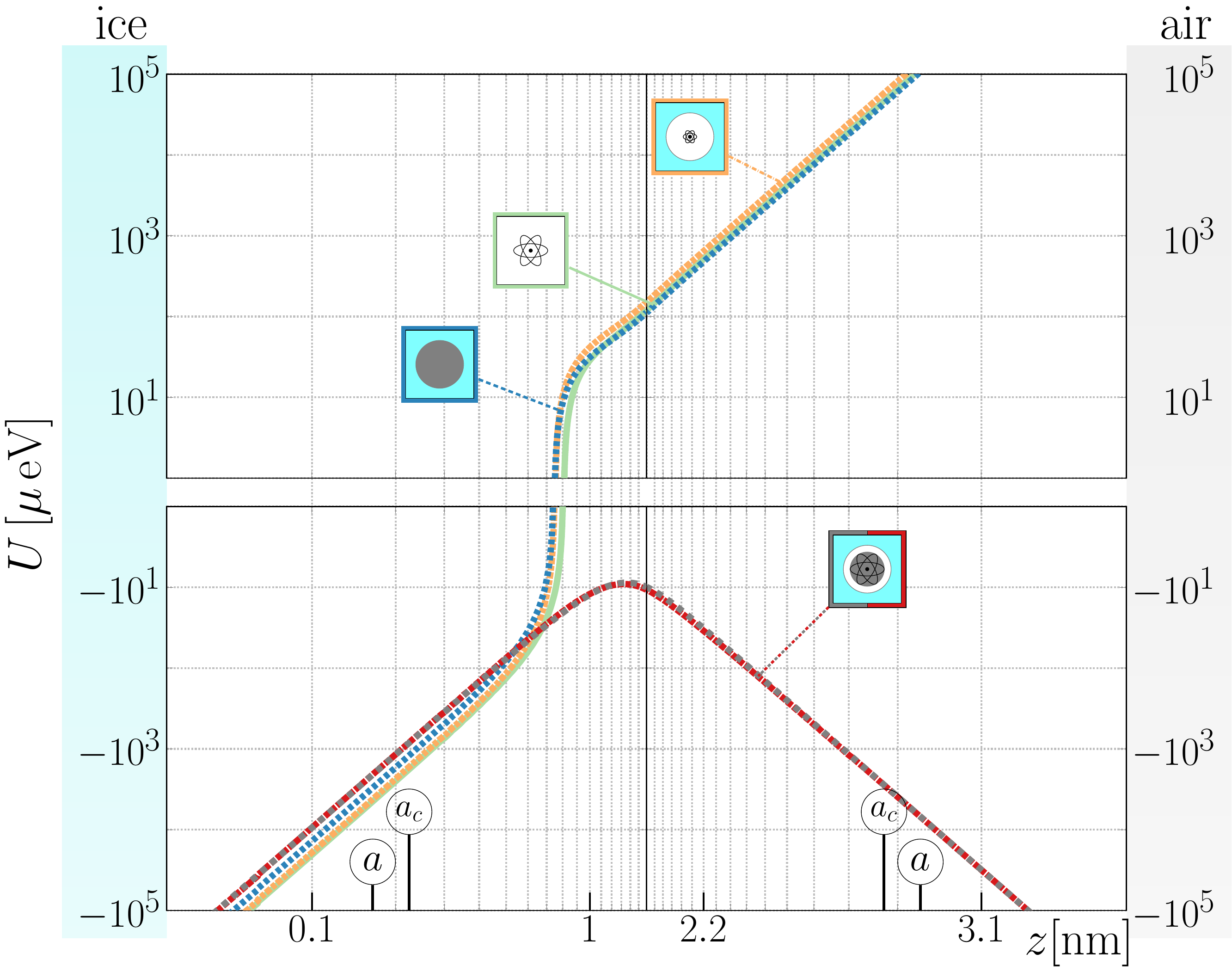}
 \caption{(Color online) Resulting Casimir--Polder potentials for carbon dioxide using the free-space polarizability (green), hard-sphere model (blue), Onsager's real cavity model (orange), the finite-size model (red) and the nonretarded limit using the finite-size model (gray). The cavity radius $a_C$ and the particles radius $a$ are marked by needles on the $x$-axis to illustrate the physically possible positions of the particles.}\label{fig:pot3}
\end{figure}

The resulting forces are attractive for the water-ice interface for all effective polarizability models for both molecules. Molecules tend to stick to the ice surface. Near the water-air interface, the simpler and more commonly used models (hard-sphere and Onsager's real cavity models) predict both molecules to be pushed away from the interface. However, for the generalized effective polarizability model the CO$_2$ is attracted to the water-air interface in sharp contrast to CH$_4$. This can be traced to the behavior of the effective polarizability being negative for a range of frequencies due to the cavity surrounding the CO$_2$ molecule inside water. These results are shown in more detail in Figs.~\ref{fig:pot2} and ~\ref{fig:pot3}. The CH$_4$ molecule residing in the water region is predicted to be attracted towards the ice surface and pushed away from the optically thinner air. A more surprising result is found for CO$_2$ where the potentially most realistic model taking into account the finite size of a particle, predicts that CO$_2$ molecules can be attracted towards the water-air interface. This predicted behavior provides a fertile testbed for experiments comparing the effective polarizability models.

An experimental test for or against the different models can be performed by a horizontal arrangement of the described three layer system that the liquid layer stays stable due to gravitational forces, similar to the experiments reported in Ref.~\cite{Enami13}. The use of solved gases in pure water and the measurement of the fraction of escaped gases will result in the solubility which follows Henry's law \cite{Henry}. Caused by the separability of the forces with respect to both interfaces, such an experiment is not limited in the height of the water layer and the choice of the supporting material as long as the condition of a low reflectivity at this interface is satisfied.

\section{Conclusions}

 While the free space polarizabilities of methane and carbon dioxide can be seen to behave in very similar ways, 
 the effective polarizabilities in water are very distinct. 
 In order to study this effect a new model has been explored that accounts for the finite size of the gas molecule in a cavity. The dimensions of the cavity are determined from the positions of the surrounding water molecules around the gas molecule as described in Ref.~\cite{JohannesJCPA2017}. This leads to the surprising conclusion that some molecules, like carbon dioxide, may potentially be pushed towards  an optically thinner region by Casimir--Polder forces. Specifically, the gas molecule in water is attracted towards air. This suggests that some greenhouse molecules may be able to  escape from melting ice more easily and enter the surrounding air. The calculations may be further refined by using a continuous cavity profile as informed by microscopic simulations of the environmental medium.


\begin{acknowledgements}
We acknowledge support from the Research Council of Norway (Projects 250346 and 243642). We thank the Australian National Computer Infrastructure (NCI).
We gratefully acknowledge support by the German Research Council (grant BU1803/3-1, S.Y.B. and J.F.) the Research Innovation Fund by the University of Freiburg (S.Y.B., J.F. and M.W.) and the Freiburg Institute for
Advanced Studies (S.Y.B.).  
\end{acknowledgements}


\begin{thebibliography}{10}
\bibitem{Mahanty2} J. Mahanty and B. W. Ninham, {\it Dispersion Forces} (Academic, London, 1976).
\bibitem{Pars} V. A. Parsegian, {\it Van der Waals forces: A handbook for biologists, chemists, engineers, and physicists}, (Cambridge University Press, New York, 2006). 
\bibitem{Ninhb} B. W. Ninham and P. Lo Nostro, {\it Molecular Forces and Self
 Assembly in Colloid}, in Nano Sciences and Biology, (Cambridge University Press, Cambridge, 2010).
\bibitem{Buhmann12a} S. Y. Buhmann, \textit{Dispersion Forces
I: Macroscopic Quantum Electrodynamics and Ground-State Casimir, Casimir--Polder and van der Waals Forces} (Springer, Heidelberg, 2012),

S. Y. Buhmann, \textit{Dispersion Forces II: Many-Body Effects, Excited Atoms, Finite Temperature and Quantum
Friction} (Springer, Heidelberg, 2012).
\bibitem{Elbaum} M. Elbaum and M. Schick, Phys. Rev. Lett. {\bf 66}, 1713 (1991).

\bibitem{Elbaum2} M. Elbaum and M. Schick, J. Phys. I France {\bf 1}, 1665 (1991).
\bibitem{Wilen} L. A. Wilen, J. S. Wettlaufer, M. Elbaum, and M. Schick, Phys. Rev. B {\bf 52}, 12426 (1995).
\
\bibitem{Osborn} S. G. Osborn, A. Vengosh, N. R. Warner, R. B. Jackson, Proc. Nat. Acad. Sci. {\bf 108}, 8172 (2011).
\bibitem{Vidic} R. D. Vidic, S. L. Brantley, J. M. Vandenbossche, D. Yoxtheimer, J. D. Abad, Science {\bf 340}, 1235009 (2013).
\bibitem{Kort} E. A. Kort, et al., Nature Geoscience {\bf 5}, 318 (2012).
\bibitem{Whiteman} G. Whiteman, C. Hope, and P. Wadhams, Nature {\bf 499}, 401 (2013).

\bibitem{JohannesJCPA2017} J. Fiedler, P. Thiyam, A. Kurumbail, F. A. Burger, M. Walter,  C. Persson, I. Brevik, D. F. Parsons, M. Bostr{\"o}m, S. Y. Buhmann, J. Phys. Chem. A {\bf{121}}, 9742 (2017).
\bibitem{PriyaPRE14}P. Thiyam, C. Persson, B. E. Sernelius, D. F. Parsons, A. Malthe-S{\o}renssen and M. Bostr\"{o}m, \textit{Phys. Rev. E} {\textbf {90}}, 032122 (2014).
\bibitem{Sambale09} A. Sambale, D.-G. Welsch, Ho Trung Dung, S. Y. Buhmann, Phys. Rev. A {\bf 79}, 022903 (2009).
\bibitem{Casimir}  H. B. G. Casimir and D. Polder. Phys. Rev. {\bf 73}, 360 (1948).
\bibitem{Buhmann} S. Y. Buhmann, D.-G. Welsch, and T. Kampf, Phys. Rev. {\bf 72}, 032112 (2005).
\bibitem{Sernelius} Bo E. Sernelius, {\it Surface Modes in Physics} (Wiley-VCH, Berlin, 2001).
\bibitem{ParsonsNinham2009} D.F. Parsons and B. W. Ninham, J. Phys. Chem. A {\bf 113}, 1141 (2009).
\bibitem{ParsonsNinham2010dynpol} D.F. Parsons and B. W. Ninham, Langmuir {\bf 26}, 1816 (2010).
\bibitem{Onsager} L. Onsager, J. Am. Chem. Soc. \textbf{58}, 1486 (1936).
\bibitem{Sambale07} A. Sambale, S. Y. Buhmann, D.-G. Welsch and M. S. Tomas, Phys. Rev. A \textbf{75}, 042109 (2007).
\bibitem{Jackson} J. D. Jackson, \textit{Classical Electrodynamics} (Wiley, New York, 1998).
\bibitem{Sambale10} A. Sambale, S. Y. Buhmann and S. Scheel, Phys. Rev. A \textbf{81}, 012509 (2010).

\bibitem{Henry}R. Sander, Atmos. Chem. Phys. \textbf{15}, 4399 (2015).
\bibitem{Enami13}S. Enami and A.J. Colussi, J. Chem. Phys. \textbf{138}, 184706 (2013).

\end{thebibliography}
\end{document}